**The Effects of Smartphones on Well-Being:**
**Theoretical Integration and Research Agenda**

Kostadin Kushlev* & Matthew R Leitao

**Georgetown University**

**\*Corresponding author**
Department of Psychology
Georgetown University
37th and O Streets, N.W.
Washington DC 20057
USA
(202) 687-3034
Kostadin.Kushlev@georgetown.edu

**Abstract.** As smartphones become ever more integrated in people's lives, a burgeoning new area of research has emerged on their well-being effects. We propose that disparate strands of research and apparently contradictory findings can be integrated under three basic hypotheses, positing that smartphones influence well-being by (1) replacing other activities (*displacement hypothesis*), (2) interfering with concurrent activities (*interference hypothesis*), and (3) affording access to information and activities that would otherwise be unavailable (*complementarity hypothesis*). Using this framework, we highlight methodological issues and go beyond net effects to examine how and when phones boost versus hurt well-being. We examine both psychological and contextual mediators and moderators of the effects, thus outlining an agenda for future research.

As smartphones become ever more ubiquitous and integrated in our lives [1], their effects on well-being have become the subject of countless opinion columns and a burgeoning new area of research. We propose that researchers' contradictory conclusions about the well-being effects of phones are largely due to methodological issues and can be integrated under an overarching theoretical framework by considering when, how, and why phones promote versus hurt well-being. Our review focuses on research published since 2018, though we draw on earlier research when necessary to provide evidence for key mechanisms and moderating factors.

## 1. Net Effects of Screen Time: The Great Debate

American adults spent an estimated 3.5 hours a day on their mobile screens in 2019, surpassing even the notoriously long time they spend in front of their TV screens [2]. In the same year, American teenagers spent more than seven hours a day in front of a screen across all devices; smartphones—the most used mobile devices by far—accounted for about half of that time [3]. Across large representative samples of adolescents, researchers have consistently found that more self-reported screen time—on all devices, digital devices only, and phones more specifically—predicts slightly lower well-being [4*,5,6*,7,8]. Importantly, these studies also show that this net negative effect is small and that the relationship is not linear: People who spend some time in front of screens feel better than those who spend no time, but heavy users feel notably worse [5, 6*].

Despite the consistency of findings, researchers disagree about their implications. Some suggest that the effect of screen time on well-being is as negligible as that of wearing glasses or eating potatoes [4*]. Others suggest that phone screen time can explain recent upswings in depression and suicide that have coincided with the increased adoption of smartphones [6*,7,8]. We suggest that we cannot draw either of those conclusions based solely on correlational analyses of self-reported screen time. People's subjective estimates of their average screen time share as little as 10% variance with their actual screen time when measured objectively [9]. Furthermore, light users tend to overestimate their screen time, whereas heavy users tend to underestimate their screen time [9]. Such measurement error, which depends on the construct measured, may change both the size of the effects and the shape of the relationship. To quantify the net effect of screen time on well-being, therefore, researchers need to begin to examine the relationship between screen time and well-being using more precise measures, such as experience sampling [10] or objective measures from usage-tracking apps [11,12**,13**,14].

## 2. It Is Not All About Screen Time: From Displacement to Interference

Even if we measured phone screen time more precisely, however, we would still get an incomplete picture of the effects of smartphones on well-being. People check their phones frequently, an estimated 50 times a day, but they only spend a little over a minute on their phones at a time with only 5% of instances lasting longer than 10 minutes [15]. Thus, though positively related, frequency and time of phone use are distinct variables, estimated to share only about 10%-15% variance [12**,13**].

According to the *displacement hypothesis* [16], screen time—regardless of whether it is on a desktop, TV, or phone—influences psychological outcomes by replacing other activities (c.f., *Bowling Alone*; [17]). From this perspective, smartphones can negatively impact well-being primarily by reducing the time we spend doing other activities that are essential for well-being, such as sleeping [18,19,20] or socializing [21]. Unlike a desktop computer, a TV set, or even a tablet, however, smartphones are within easy reach throughout the day, often while we are engaging in other activities: from exercising and sleeping to working, socializing, and watching TV [3,22,23]. In other words, smartphones are distinct from other, similarly versatile computing devices by virtue of their portability. Thus, even when phone use is not replacing other activities, phones can still interfere with concurrent activities (Table 1). According to the *interference hypothesis* [24,25], the frequency of phone use is critical for understanding the effects of smartphones on well-being (Table 1). During lunch with a friend, for example, we might spend only a small fraction of the time looking at our phone screen, but our conversation could still be disrupted by brief yet frequent phone checks.

### 2.1. Attention

The displacement and interference hypotheses both assume that the net effects of phones on well-being, at any given time, will depend on the well-being people gain by using their phones, minus the well-being that people would have gained had they not used their phones. The two hypotheses, however, make unique, though not mutually exclusive, predictions about the role attention has in explaining the effects of smartphones (Table 1). Since attention plays a critical role in reaping positive emotions from positive experiences [26,27], screen time should produce greater benefits for well-being as people become more attentive to their screens. Thus, the displacement hypothesis suggests that attention should moderate the relationship between screen time and well-being. Greater immersion in video gaming, for example, can magnify the emotional effects of gaming [28]. In contrast, the interference hypothesis predicts that phones will influence well-being to the extent that they distract people from other concurrent activities. In this case, attention should act as a mediator of the effects of phones on well-being. Indeed, across a variety of social situations—from spending time with one's children [29] to sharing a meal with friends [30]—phones decrease well-being precisely by fragmenting people's attention [24].

## 2.2. Notifications

Unlike the displacement hypothesis, the interference hypothesis suggests that phones can impact well-being even when they are not being used as phones can also fragment attention exogenously through alerts and notifications. According to the interference hypothesis, therefore, reducing exogenous phone interruptions should improve well-being. Indeed, a 14-day field experiment found that receiving phone notifications in three daily batches, compared to receiving them as usual, made users happier and less stressed [31*]. Consistent with the interference hypothesis, these effects of batching notifications on well-being were mediated by an improvement in participants' subjective quality of attention.

## 2.3. Mere presence

Recent research has suggested that even the mere presence of one's smartphone can impair basic cognitive processes [32,33]. Having one's phone within easy reach does not directly impair people's ability to sustain attention, but it does impair cognitive capacity, such as working memory [33]. Based on the interference hypothesis, we can predict that the mere presence of one's phone should only interfere with one's ability to enjoy cognitively demanding activities (e.g., a deep conversation), but not simpler pleasures (e.g., a beautiful sunset). Consistent with these predictions, a small study found that the mere presence of a mobile phone during a dyadic social interaction, prevented people from cultivating a sense of social connection, trust, and empathy when discussing a meaningful topic but not when discussing a more casual topic: plastic holiday trees [34]. Though a recent study failed to replicate these findings [35], both studies were likely underpowered ($N$s < 100) to detect an interaction effect from a subtle manipulation [36]. Thus, the effect of mere presence on well-being deserves further investigation.

### 3. Beyond Net Effects: Context and Affordances

Across screen time, frequency of use, and the mere presence of phones, we have so far seen that phones have small but consistently negative, net effects on well-being. Do the devices that allow us to deposit a check from the comfort of our homes and check in with faraway family and friends have no positive effect on our well-being?

The displacement and interference hypotheses do not predict that phone use will always decrease well-being. Rather, phones should result in negative effects on well-being only when they displace or interfere with activities essential for well-being (e.g., sleep) and with nondigital experiences that afford a greater source of well-being than their digital counterparts [37]. For example, receiving social support in-person produces greater benefits for well-being than receiving support via text message [38]. Still, the effects of phones in any given situation should depend on the affordances and limitations of a person's current environment [39]. Indeed, although phone-mediated communication, such as text messaging, predicts lower well-being outcomes when people are concurrently socializing face-to-face, phone-mediated communication has little effect on well-being when people are alone [40].

#### 3.1. Complementarity

While not as impactful as talking with somebody in person, receiving social support via text message is better for well-being than receiving no support at all [38]. Thus, we must formulate a final *complementarity hypothesis*: Smartphones can boost well-being by affording access to information, communication, and experiences that would otherwise be unavailable. When people have to locate an unfamiliar building, for example, those who can use their phones to navigate feel happier than those who have to use more old-fashioned methods, such as asking locals for directions [41]. When people are in an otherwise stressful situation, phone use provides a level of comfort that is unattainable through their less portable and personal counterparts, laptops [42**]. And after experiencing social rejection, even the mere presence of one's phone can be a source of comfort—as assessed by both self-report measures and biological markers of stress [43*]. Smartphones can and do provide benefits for well-being, but those benefits depend on people's current emotional experience, the contextual affordances, and the relevance of their phone use to current goals and activities (Table 1). In this context, it is hardly surprising that the net effects of phone screen time [4*] or frequency of use [13**] are small to negligible.

## 4. Conclusion and Future Research Agenda

Although the first iPhone was introduced only a dozen years ago, the effects of smartphones on well-being have become a burgeoning new area of research. Based on the existing findings, we can draw three (very) preliminary conclusions:

1) Smartphones can either boost or hurt well-being depending on when and how they are used;

2) Phones seem to have a net negative effect on well-being overall

3) The net effects of phones on well-being are typically small to negligible but become substantial with heavy, problematic (over)use (i.e., use that is likely to displace or interfere with other activities).

Though it is possible to draw these preliminary conclusions, several persistent methodological issues need to be addressed. First, correlational findings need to be tested using experimental methods in order to draw causal conclusions. Second, researchers need to pay more attention to measurement (1) by distinguishing between screen time, frequency of use and notifications, and mere presence, and (2) by moving beyond imprecise and biased self-report measures of phone use to more objective measures obtained over multiple days through mobile sensing [11,12**,13**,14,44]. Third, research needs to focus more on understanding when these multifunctional devices are beneficial versus detrimental to well-being.

We have proposed an overarching *displacement-interference-complementarity* (D.I.C.) theoretical framework (Table 1), which integrates apparently contradictory findings by considering how, when, and why phones impact well-being. In particular, this framework considers the role of both basic psychological processes and contextual factors as mediators and moderators. Importantly, the framework allows for the integration of new findings, processes, and interactions. For example, recent research on phubbing—snubbing others through phone use [25]—has focused on the interpersonal effects of phones. Research in this area has suggested that phubbing can negatively impact others across interactions between friends [45], romantic partners [46,47], and parents and children [48,49,50]. Within our framework, these findings can be understood as interference effects mediated by the quality of the user's attention as perceived by others.

Finally, because the D.I.C. framework is formulated around the distinguishing factor of smartphones—their portability—the framework can easily be applied to the increasingly portable technologies of the future. Only through such theory-driven approaches, can we ensure that today's research on the psychological effects of smartphones will transcend current technology—and contribute to our broader understanding of the psychological effects of humanity's evolving relationship with technology.

## Acknowledgements

The authors thank Maureen Tibbetts and Amanda Van Orden for helping to polish the article for publication.

The authors used three nationally representative samples of adolescents (N = 221,096) from the United States and the United Kingdom. Based on self-reported measures of screen time, the authors formed single composites of total digital screen time across all media and devices (excluding TV sets). Light users—adolescents spending an hour or less of screen time a day—reported the highest well-being, with higher use predicting lower well-being. Heavy users, those who spent 5 or more hours a day on a digital screen—experienced the biggest detriments to well-being, including being twice as likely as light users to have attempted suicide.

In this two-week field experiment, participants whose notifications were pushed in three daily batches felt happier and less stressed than control participants who received their notifications as usual. Participants in a third condition who got no notification alerts did not feel more attentive than controls because they self-initiated more phone checking. In a fourth condition, batching notifications once an hour did not have detectable effects on these outcomes compared to control.

Table 1. *The Displacement-Interference-Complementary Framework for the effects of smartphones on well-being.*

| | Core Prediction | Primary predictors | Mediators/Mechanisms | Moderators/Interactions |
|---|---|---|---|---|
| | | What phone-mediated behaviors to measure? | How do phones affect well-being? | When do phones predict higher vs. lower well-being? |
| **Displacement Hypothesis** | Phones influence well-being by replacing other activities. | Time spent on phone-mediated activities (screen time). | Time spent on other activities (e.g., sleep, face-to-face interactions). | 1. Quality of phone-mediated activities.<br>2. Quality of attention (immersion).<br>3. Well-being affordances of activities otherwise available. |
| **Interference Hypothesis** | Phones influence well-being by interfering with concurrent activities. | 1. Frequency of phone checking, notifications.<br><br>2. Mere presence of phones. | 1. Quality of attention.<br><br>2. Cognitive capacity. | 1. Well-being affordances of concurrent experiences.<br><br>2. Cognitive demand of concurrent activities (low vs. high). |
| **Complementarity Hypothesis** | Phones influence well-being by affording information and activities not otherwise available. | Type of phone use: information communication entertainment | 1. Efficiency (time saved).<br>2. Buffer negative emotions.<br>3. Source of positive emotions. | 1. Availability of unmediated sources of information, communication, and entertainment.<br>2. Relevance of phone use to concurrent activities, goals, and needs.<br>3. Quality of current experience (positive, neutral, negative). |